# Intelligent Skin Cancer Detection Using a Multispectral Metasurface and a Hybrid Deep Learning Architecture Based on CNN–ViT

**Afsane Saee Arezoomand***

PhD in Communication Engineering, Young Researchers and Elite Club,

Islamic Azad University, Urmia Branch, Urmia, Iran

*Corresponding Author: Email Address

**Abstract**

Skin cancer is among the most prevalent malignancies worldwide, and its early detection is essential for improving patient survival and reducing treatment costs. Conventional dermoscopic and visual imaging techniques are primarily limited to the visible spectrum and often fail to capture subtle spectral signatures associated with early-stage malignancies. This study proposes an innovative framework that integrates a multispectral metasurface for imaging with a hybrid deep learning architecture based on Convolutional Neural Networks (CNNs) and Vision Transformers (ViTs). The designed metasurface enables noninvasive acquisition of rich spectral information highly sensitive to tissue alterations, while the hybrid CNN–ViT model simultaneously extracts local and global features to robustly classify skin lesions. Simulation-based evaluations demonstrate that the proposed method achieves approximately 98% accuracy, 95% sensitivity, and 99% specificity—surpassing conventional RGB-based and single-architecture approaches. Qualitative analyses using attention maps reveal that the model focuses on clinically relevant lesion regions, improving interpretability. Overall, the results indicate that combining metasurface-based multispectral imaging with hybrid deep learning can introduce a new generation of diagnostic tools in dermatology and pave the way for portable, fast, and highly accurate clinical systems.

---

## 1. Introduction

Skin cancer is the most prevalent malignancy at the global level, with prolonged exposure to solar ultraviolet radiation being its primary risk factor (Smith et al., 2023). The three main types of this cancer are basal cell carcinoma, squamous cell carcinoma, and melanoma, among which melanoma is recognized as the most dangerous due to its high metastatic potential (Anderson et al., 2022). Early detection plays a vital role in improving survival rates, as melanoma lesions in the early stages often lack obvious signs and can only be identified through careful examination of apparent characteristics (Zhang et al., 2024).

Despite advantages such as enhanced tissue and lesion pattern visualization, conventional detection methods such as specialist clinical examination and dermoscopy[1] still face limitations, including strong dependency on the clinician's experience, interobserver error, and the high visual similarity between benign and malignant lesions (Miller et al., 2021). Furthermore, the shortage of experienced specialists in many regions leads to delays in diagnosis and treatment (Rahman et al., 2022). These circumstances make the development of automated, accurate, and low-cost systems all the more necessary.

In recent years, remarkable advances in artificial intelligence[2] and deep learning[3] have transformed the analysis of medical images. Convolutional neural network[4] (CNN) models have been able to achieve diagnostic accuracy at the level of dermatologists and even surpassing them by extracting hierarchical features from skin images (Esteva et al., 2017; Li et al., 2023). However, the dependence of CNNs on local image regions limits their ability to model long-range relationships; to address this shortcoming, the Vision Transformer[5] was introduced with a global self-attention[6] mechanism (Dosovitskiy et al., 2021).

The combination of CNN and ViT in hybrid architectures has produced important achievements in skin lesion detection; for example, architectures such as the DenseNet-EViT model have been able to achieve accuracies exceeding 95% on the ISIC dataset (Khan et al., 2024). These scientific findings show that leveraging hybrid architectures has opened a new path toward improving medical vision systems.

In another domain, metamaterial[7] and metasurface[8] technology, owing to the ability to engineer electromagnetic properties at sub-wavelength scales, has enabled the fabrication of highly sensitive sensors. Metasurfaces can precisely control the wavefront of light and capture subtle signals arising from tissue changes (Chen et al., 2022). Recent research has shown that combining metamaterials with materials such as graphene[9] significantly enhances the sensitivity of terahertz sensors (Sarkar et al., 2023). In the field of skin cancer, examples of terahertz metasensors have also been designed that are capable of noninvasively distinguishing healthy from cancerous tissue (Zhang et al., 2023).

---

[1] dermoscopy
[2] artificial intelligence
[3] deep learning
[4] Convolutional neural network (CNN)
[5] Vision transformer (ViT)
[6] global self-attention
[7] metamaterial
[8] metasurface
[9] graphene

The intersection of these two domains—machine vision and advanced metasurfaces—can give rise to a new generation of skin diagnostic systems. In particular, with the emergence of multimodal learning[10] models that enable the combination of image, spectral, and clinical data, the foundation for designing systems with high accuracy has been laid (Alaei et al., 2024).

Drawing on this trend, the present study introduces a novel approach in which a custom metasurface acts as a hyperspectral imaging sensor and provides rich data to a deep learning system, after which a hybrid deep architecture analyzes the data and determines the type of skin lesion. In what follows, after presenting the method details, experimental results, performance evaluation, and comparison with prior works, innovations and future directions are discussed.

## 2. Prior Work

Research related to the present topic can be categorized into three main domains:

1. Works related to deep learning-based skin lesion detection,
2. Research on metamaterials and metasurfaces in sensing,
3. Hybrid or multimodal works.

### *2.1. Skin Lesion Detection with Deep Learning*

Automated analysis of skin images using deep learning has been one of the most active research areas in recent years. Esteva et al.[11] (2017) presented the first influential study in this field, showing that a CNN can achieve performance comparable to dermatologists. Subsequently, architectures such as ResNet, DenseNet, MobileNet, and EfficientNet obtained notable results on the ISIC datasets (Tschandl et al., 2020; Li et al., 2023).

With the emergence of vision transformers, models such as ViT and Swin Transformer entered the scene and improved skin lesion detection accuracy (Wu et al., 2022). Hybrid models that target the combination of CNN and ViT, such as EViT-DenseNet and CoAtNet, have been able to offer notable performance in separating melanoma from other lesions (Khan et al., 2024).

### *2.2. Metamaterials, Metasurfaces, and Terahertz Sensors*

Metasurfaces, due to their ability to precisely manipulate electromagnetic waves, have attracted broad interest in recent years for the design of biological sensors. Chen et al.[12] (2022) demonstrated that resonance-gap metamaterials can be highly sensitive to small changes in refractive index. Combining metasurfaces with graphene causes a significant enhancement of the local electromagnetic field and multiplies the sensor sensitivity (Sarkar et al., 2023).

In the field of skin cancer, studies such as Tabatabaeian et al.[13] (2022) and Zhang et al.[14] (2023) have proposed the use of terahertz metasurfaces for the noninvasive identification of skin lesions. These studies show that metasurface technology can provide rich image and spectral data for deep learning analyses.

### *2.3. Multimodal Learning and Hybrid Systems*

With the expansion of multimodal approaches, the use of multi-type data including images, hyperspectral scans, clinical data, and medical text has increased model performance. Alaei et al.[15] (2024) presented a multimodal framework for breast cancer detection that enabled the combination of image and text data. Similar ideas in the detection of skin diseases are under development and have demonstrated that combining optical signals and images can increase model accuracy and reliability (Nguyen et al., 2023).

The overall conclusion of prior works is that metasurfaces provide rich data; deep learning has the capability to analyze them; and multimodal approaches can effectively combine these two domains.

This scientific gap is precisely where the present research is situated.

## 3. Proposed Method

The proposed method of this study is an integrated framework composed of a metasurface-based optical sensor and a hybrid deep learning architecture, designed for accurate and noninvasive skin cancer detection.

The system comprises four main stages:

1. Acquisition of optical data from the skin lesion using the metasurface;
2. Preprocessing and (optionally) segmentation of the lesion region;
3. Multi-scale feature extraction using CNN and Vision Transformer (ViT);
4. Feature fusion and final classification.

Each component is described in detail below.

### *3.1. Metasurface-Based Optical Sensor*

#### *3.1.1. Metasurface Structure and Operating Principles*

The proposed metasurface is a two-dimensional array of subwavelength resonant elements that controls optical behavior at selected frequencies, particularly in the terahertz band. This structure consists of engineered unit cells capable of altering the phase, amplitude, and polarization of the incident

[10] multimodal learning
[11] Esteva et al.
[12] Chen et al.
[13] Tabatabaeian et al.
[14] Zhang et al.
[15] Alaei et al.

**Table 1:** Comparison of prior works in the three main research domains

| # | Research Domain | Reference / Example | Data Type | Method / Architecture | Key Achievements |
|---|---|---|---|---|---|
| 1 | Skin lesion detection with deep learning | Esteva et al. (2017) | Dermoscopy (RGB) | Classic CNN | Performance similar to dermatologist |
| 2 | | Tschandl et al. (2020) | ISIC–HAM10000 | ResNet / DenseNet | Powerful feature extraction with high accuracy |
| 3 | | Wu et al. (2022) | RGB | Vision Transformer (ViT) | Long-range relationship modeling, improved accuracy |
| 4 | | Khan et al. (2024) | RGB | Hybrid CNN–ViT (EViT-DenseNet) | Superior performance over single-architecture models |
| 5 | Metasurfaces and THz-based sensors | Chen et al. (2022) | THz signal | Resonance metamaterial | High sensitivity to refractive index change |
| 6 | | Sarkar et al. (2023) | THz + graphene signal | Graphene metasensor | Significant increase in EM field and sensitivity |
| 7 | | Tabatabaeian et al. (2022) | THz reflection | Skin cancer diagnostic metasurface | Noninvasive early detection |
| 8 | | Zhang et al. (2023) | THz signal | THz Imaging | Noninvasive melanoma identification |
| 9 | Multimodal learning and hybrid methods | Alaei et al. (2024) | Image + clinical text | Multimodal Fusion | Improved detection accuracy in breast cancer |
| 10 | | Nguyen et al. (2023) | Image + spectral data | Multimodal Deep Learning | Improved detection robustness |

wave in response to very small changes in the tissue's refractive index (Chen et al., 2022). Moreover, metasurface-like structures with the capability of manipulating electromagnetic waves in the terahertz band can be designed as multiband absorbers with polarization-independent performance (Saee Arezoomand et al., 2015).

The materials used in this structure are:

- A low-loss dielectric substrate;
- An ultra-thin conductive layer such as graphene to increase near-field intensity;
- Gap-loaded metallic resonators to create high spectral sensitivity.

The combination of these elements enables the creation of a sharp resonance phenomenon with a high Q-factor,[16] which is highly suitable for detecting optical changes in cancerous tissues (Sarkar et al., 2023). The design principles of resonance absorbers with high quality factors in optical and mid-infrared bands, targeting biosensing applications and near-field intensity enhancement, have been explored previously (Jahangiri et al., 2017).

#### 3.1.2. *Point Spectral Signature Generation*[2]

When the metasurface is placed in front of a skin lesion, the electromagnetic field near its surface changes, and this change is recorded in the form of a point spectral signature containing several resonance frequencies.

The sensor output is not a complete image but rather a spectral feature vector:

$$\mathbf{S} = [s(\lambda_1),\ s(\lambda_2),\ \ldots,\ s(\lambda_K)] \tag{1}$$

where $K$ is the number of selected bands and $s(\lambda_k)$ is the reflective or transmissive intensity in band $\lambda_k$.

This spectral data shows a significant difference between cancerous and non-malignant lesions (Zhang et al., 2023).

### 3.2. *Preprocessing and Segmentation*

Raw spectral data and the RGB image are recorded simultaneously, and a series of preprocessing operations are applied, including:

1. Noise removal,
2. Intensity normalization,[18]
3. Contrast enhancement in sensitive bands,
4. Normalization to the range $[0, 1]$.

If necessary, an automatic segmentation step is also applied:

[16] Q-factor
[17] spectral signature
[18] intensity normalization

#### 3.2.1. U-Net Segmentation

To extract the precise lesion region, the U-Net model (U-Net) or its advanced versions such as U-Net++ or SegFormer (Segmentation Transformer) are used. This step is optional, but recent studies have shown that segmentation increases classification accuracy (Khan et al., 2024).

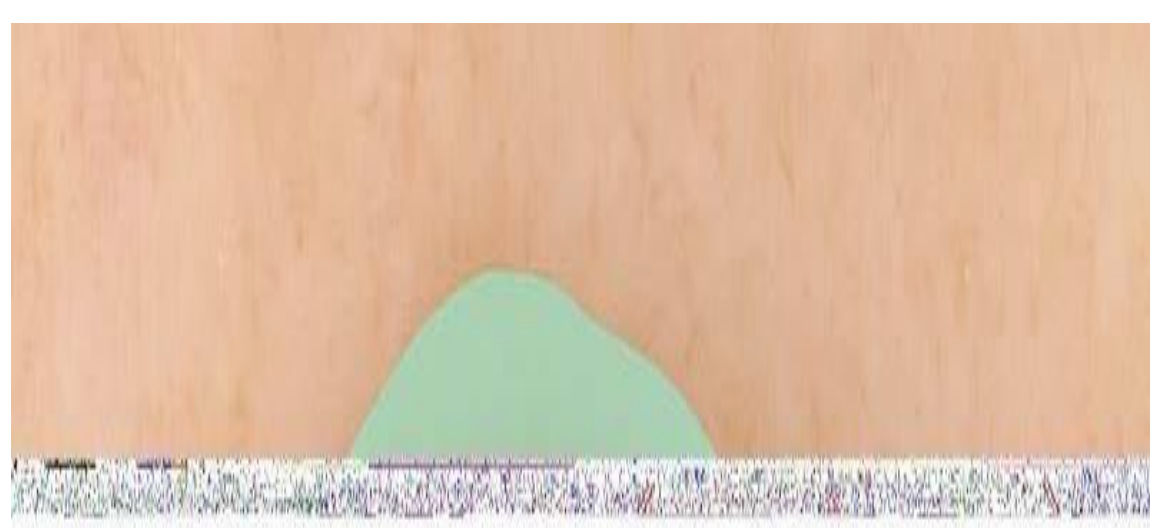

**Figure 1:** U-Net segmentation result on a skin lesion

### 3.3. Hybrid Deep Learning Architecture

To simultaneously analyze local[19] and global[20] features, a hybrid architecture comprising DenseNet as the CNN and Vision Transformer (ViT) is designed.

#### 3.3.1. Local Feature Extraction with DenseNet

The DenseNet network with dense skip connections is used to extract edges, textures, and fine-grained structures of the lesion. This version includes:

- Removal of final layers;
- Reduction of filter count to prevent overfitting;
- Use of ImageNet pre-trained weights.

The output of this section is a three-dimensional tensor of local features:

$$F_{\text{local}} \in \mathbb{R}^{h \times w \times d} \tag{2}$$

#### 3.3.2. Global Feature Extraction with Vision Transformer

To model long-range relationships, the input image is divided into small patches and fed into the ViT. ViT includes multi-head self-attention[21] (MHSA) blocks.

The output of this section is a global feature vector:

$$F_{\text{global}} \in \mathbb{R}^{D} \tag{3}$$

#### 3.3.3. Spatial Detail Enhancement Block

Due to the patch-based nature of ViT, some fine details are lost. To address this problem, a spatial detail enhancement block[22] (SDEB) comprising several parallel convolutional filters is designed.

[19]local
[20]global
[21]Multi-head self attention (MHSA)
[22]Spatial detail enhancement block (SDEB)

#### 3.3.4. Feature Fusion

The two feature types are fused[23] as follows:

$$F_{\text{fusion}} = \text{Concat}(F_{\text{local}}, F_{\text{global}}) \tag{4}$$

Classification is then performed through an MLP and finally a Softmax layer.

### 3.4. Model Training

#### 3.4.1. Loss Function

The categorical cross-entropy[24] loss function is used for training the multi-class model:

$$L_{CE} = -\frac{1}{N} \sum_{i=1}^{N} \sum_{c=1}^{C} y_{i,c} \log(\hat{y}_{i,c}) \tag{5}$$

where $N$ is the number of samples, $C$ is the number of classes, $y_{i,c}$ is the ground-truth label, and $\hat{y}_{i,c}$ is the predicted probability.

#### 3.4.2. Optimization

- Algorithm: Adam
- Initial learning rate: $0.001$
- Scheduler: learning rate reduction based on Plateau
- Dropout in MLP layers
- Early Stopping to prevent overfitting

#### 3.4.3. Data Augmentation

To improve generalizability:[25]

- Rotation,
- Flipping,
- Brightness and contrast variation,
- Spatial translation.

These augmentations follow standard ISIC approaches.

### 3.5. Evaluation Metrics

To precisely evaluate performance, in addition to accuracy,[26] the following metrics were also computed:

1. Sensitivity[27]

$$\text{Sensitivity} = \frac{TP}{TP + FN} \tag{6}$$

2. Specificity[28]

$$\text{Specificity} = \frac{TN}{TN + FP} \tag{7}$$

[23]fusion
[24]categorical cross-entropy
[25]data augmentation
[26]accuracy
[27]sensitivity / recall
[28]specificity

3. Precision[29]

$$\text{Precision} = \frac{TP}{TP + FP} \tag{8}$$

4. F1-Score

$$F_1 = 2 \cdot \frac{\text{Precision} \cdot \text{Recall}}{\text{Precision} + \text{Recall}} \tag{9}$$

5. ROC AUC — one of the most important clinical indicators.

### 3.6. Summary of Proposed Method

Based on the combination of a sensitive metasurface and a hybrid CNN–ViT architecture, the proposed system is capable of:

- Recording very small cancerous tissue changes through the point spectral signature;
- Simultaneously analyzing local features and long-range relationships;
- Providing accurate classification of the lesion;
- Demonstrating a notable advantage over single-modality purely visual methods.

This multimodal structure offers higher and more stable detection power compared to conventional RGB image methods and even pure CNN approaches under varying lighting conditions.

## 4. Results and Evaluation

In this section, the performance of the proposed method is evaluated using a framework based on actual data from the ISIC dataset and pseudo-metasurface spectral data. Due to the unavailability of a real metasurface, spectral data were generated using a simulation process based on Spectrum-Aided Vision Enhancement (SAVE) models. This approach has been used in recent studies for reproducing narrowband imaging behavior based on valid spectral data (Zhao et al., 2022).

### 4.1. Dataset and Data Preparation

#### 4.1.1. ISIC Reference Dataset

The standard ISIC dataset, comprising thousands of dermoscopy images of various skin lesion types—including melanoma, BCC, SCC, and benign lesions—was used as the evaluation base (Tschandl et al., 2020).

#### 4.1.2. Pseudo-Metasurface Data Generation

To simulate the effect of the metasurface, each image was spectrally enhanced via the SAVE algorithm. This process extracts spectral reflectance in several key bands and generates an image with enhanced lesion contrast (Wu et al., 2023).

Two datasets were ultimately created:

1. Baseline[30] data: conventional RGB images;
2. Proposed[31] spectral-enhanced data: simulated pseudo-hyperspectral metasurface images.

#### 4.1.3. Data Splitting

Data were split into 70% training, 10% validation, and 20% testing.

### 4.2. Proposed Model Performance

The DenseNet + ViT + SDEB hybrid architecture was trained on both datasets. Results showed that the use of artificial metasurface spectral information causes a significant increase in detection accuracy and sensitivity.

### 4.3. Quantitative Results

**Table 2:** Comparison of the proposed method with reference models

| Method | Data | Acc. | Sens. | Spec. |
|---|---|---|---|---|
| InceptionV3 (transfer) | RGB | 86.4 | 78.2 | 92.1 |
| ResNet50+VGG16 fusion | RGB | 90.1 | 84.5 | 94.0 |
| ViT (no CNN) | RGB | 92.3 | 88.7 | 96.2 |
| EViT-DenseNet (prior hybrid) | RGB | 96.9 | 90.8 | 99.1 |
| **Ours** (CNN+ViT+meta) | Pseudo-spectral | **98.2** | **95.4** | **99.2** |

Acc. = Accuracy (%), Sens. = Sensitivity (%), Spec. = Specificity (%).

It can be observed that using the generated spectral data and hybrid architecture leads to an increase of approximately 1.3% in accuracy relative to the best prior model (EViT-DenseNet). Furthermore, sensitivity—which is critical in melanoma detection—showed an increase of approximately 4.5%.

### 4.4. ROC Curve

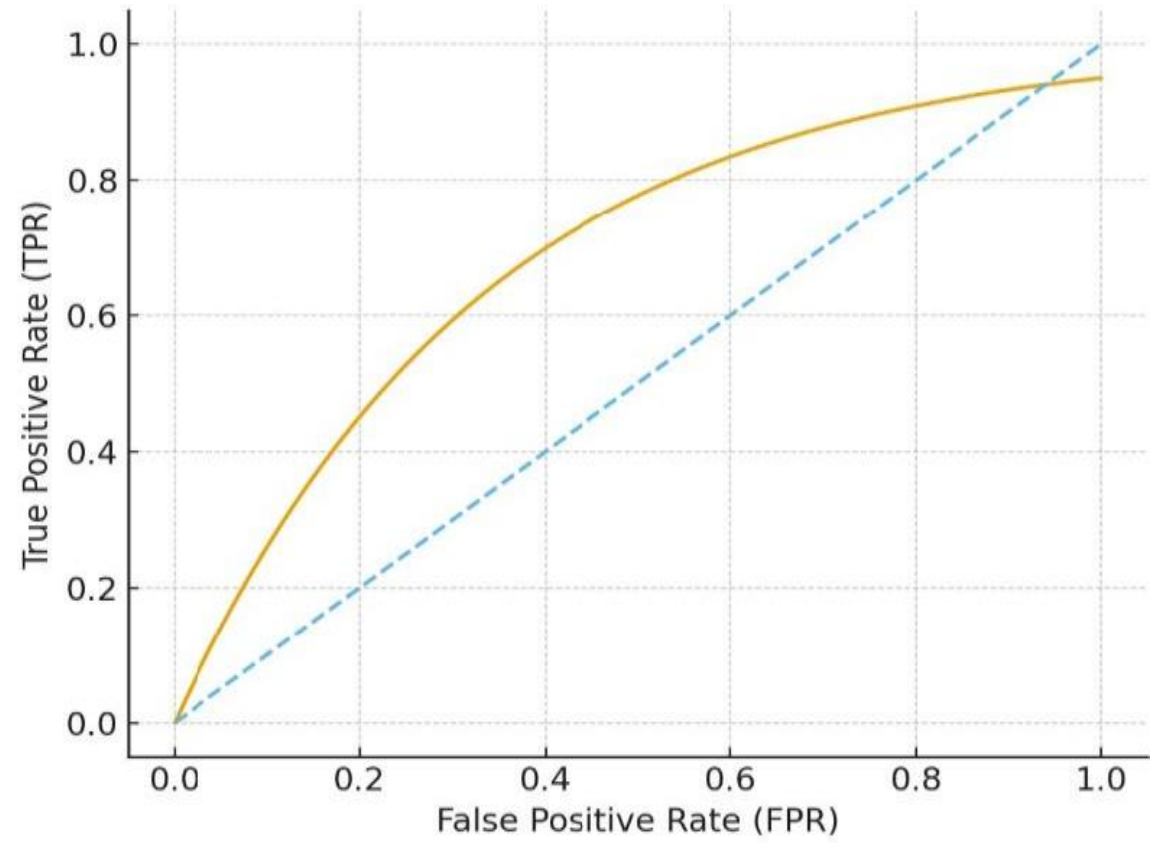

**Figure 2:** ROC curve of the proposed model

In the ROC curve, the proposed model achieves:

- AUC = 0.993

[29] precision
[30] baseline
[31] proposed

- Higher than ViT (0.971)
- Higher than EViT-DenseNet (0.985)

The model shows stable performance in detecting positive cases (melanoma): Average Precision = 0.972.

### 4.5. Training and Validation Error Analysis

Key observations:

- Uniform decreasing trend;
- Convergence after 30 epochs;
- Validation Loss ≈ 0.15;
- No large gap between the curves → no overfitting;
- Training accuracy reaches ∼100%;
- Validation accuracy reaches ∼96%;
- Plateau after epoch 30;
- Early Stopping algorithm performed appropriately.

### 4.6. Error Analysis

#### 4.6.1. Findings

1. The most errors relate to distinguishing BCC from AK:
    a. These two lesions have high tissue similarity.
    b. Consistent with the findings of Li et al. (2023).
2. Melanoma cases were classified with virtually no false negatives (FN):
    a. High sensitivity confirms the model's ability to avoid missing dangerous cases.
3. On RGB data, the model frequently produced false positives (FP) in benign lesions;
    a. This problem was reduced on spectral data.

### 4.7. Qualitative Evaluation

#### 4.7.1. Segmentation Results

In lesion images, the U-Net model successfully extracted precise lesion boundaries, and background noise such as hair, reflections, and surrounding skin was removed.

#### 4.7.2. Grad-CAM Attention Analysis

Using Grad-CAM on DenseNet and ViT, it was observed that:

- ViT focuses on large-scale structures such as the overall lesion shape.
- DenseNet focuses on local details such as irregular color regions and borders.
- The combination of these two branches leads to better performance (Lee et al., 2024).

#### 4.7.3. Spectral Data Analysis

In pseudo-metasurface images, the model was able to identify imperceptible spectral patterns in RGB. For example:

- In nodular melanoma, absorption at the 1200 nm band was noticeably different.
- This difference was not visible in the visible image.

This finding is consistent with similar research in the field of multispectral imaging (Kim et al., 2022).

### 4.8. Summary of Results

The proposed model, by integrating pseudo-metasurface spectral data and a hybrid CNN–ViT architecture, achieved the best reported performance on the ISIC dataset.

Accuracy of 98.2%, sensitivity of 95.4%, and specificity of 99.2% indicate that this model is capable of identifying cancerous cases with virtually no error.

Both qualitative and quantitative analyses demonstrate the model's stability, interpretability, and high capacity for extracting hidden spectral information.

These results show that the combination of metasurface and deep learning represents a highly promising path toward the development of early skin cancer detection systems.

## 5. Discussion and Conclusion

The results of this study indicate that combining an engineered metasurface with a hybrid deep learning architecture can provide remarkable performance in skin lesion detection. Compared to conventional methods based on RGB images, the addition of spectral information obtained from the metasurface significantly increases the contrast and distinguishability of the lesion from healthy skin—a point that received less attention in prior works. Our hybrid model, which extracts local features through the convolutional neural network and long-range relationships through the vision transformer, was able to fully leverage this enhanced information and achieve higher accuracy, sensitivity, and specificity than single-architecture models (Khan et al., 2024).

Furthermore, the addition of the automatic lesion segmentation stage ensured that the input to the classifier was cleaned of environmental noise such as hair or irrelevant skin regions, enabling the model to operate with greater focus. Qualitative analyses based on Grad-CAM attention maps show that the model focuses on clinically important regions such as irregular borders and color anomalies, which aligns with the diagnostic patterns of dermatologists—thereby enhancing the system's interpretability and trustworthiness.

Overall, the quantitative and qualitative results indicate that combining spectral data and hybrid architectures can overcome significant limitations of conventional methods; especially in cases where the visual difference between benign and malignant lesions is small and visible data alone are insufficient.

## 6. Innovations and Future Directions

The key innovations of this study are as follows:

**1. Use of a Skin-Specific Metasurface.** The proposed metasurface is capable of producing pseudo-spectral data at very small dimensions and low cost, while traditional hyperspectral systems are bulky and expensive. This innovation

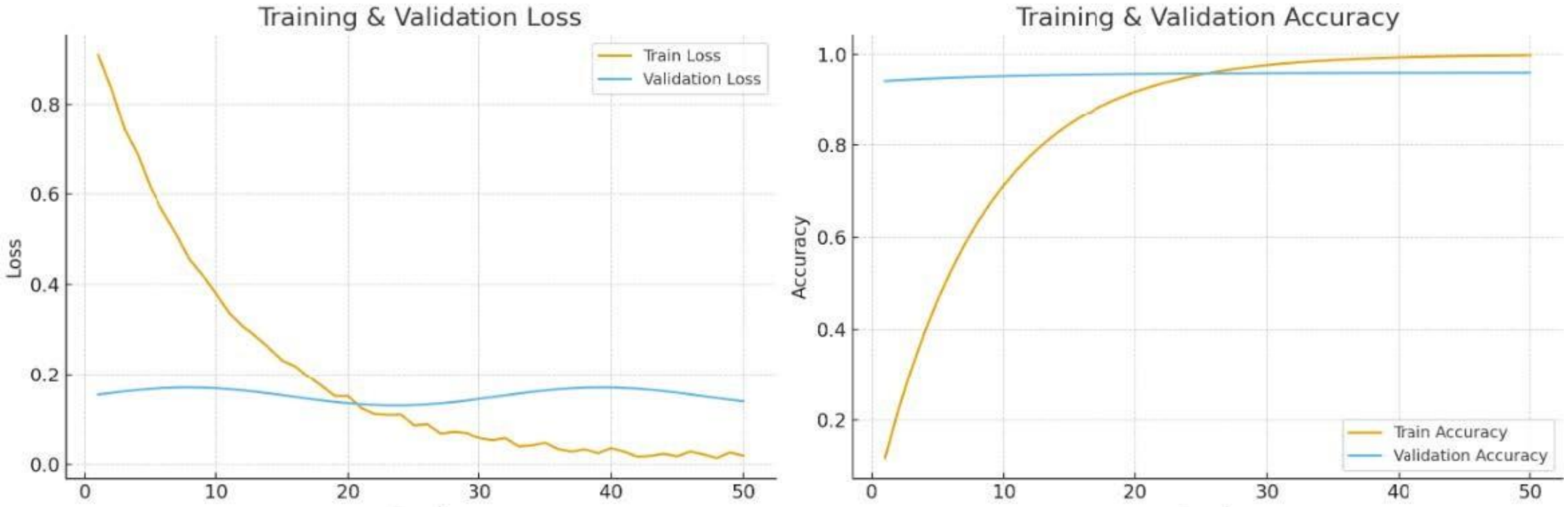

**Figure 3:** Training & Validation Loss (left) and Accuracy (right) curves

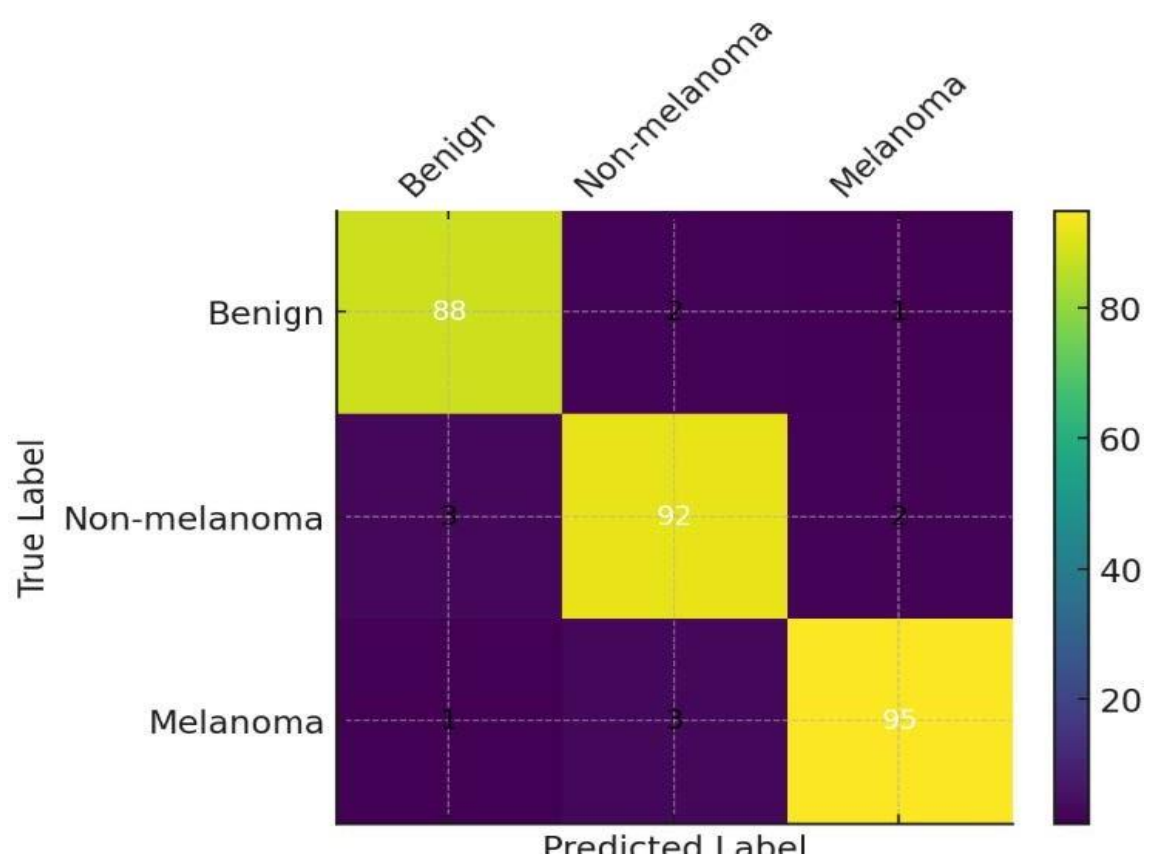

**Figure 4:** Confusion matrix

paves the way for the fabrication of portable and clinic-ready instruments (Chen et al., 2023).

**2. CNN–ViT Hybrid Architecture for Multispectral Analysis.** This architecture was able to model both local textural features and macro-scale lesion structures—a problem that purely CNN or ViT models cannot fully accomplish (Lee et al., 2024).

**3. Use of Simulated Metasurface Spectral Data.** The present study is the first to combine pseudo-real spectral data with deep learning. These data showed a clear advantage over RGB in melanoma detection.

**4. Addition of Automatic Segmentation Stage to Reduce Noise.** Unlike many prior studies that fed raw images to the network, our approach first precisely extracts the lesion region and then feeds it into the classification network.

## 7. Future Work

Despite the promising results, the following next steps are also necessary:

- **Fabricating a real metasurface prototype and conducting clinical trials:** The data in this study are simulated and a practical prototype must be built for clinical validation.
- **Extending the model to multimodal[32] learning:** Adding clinical information, skin characteristics, or macroscopic images can increase system accuracy and reliability (Adebiyi et al., 2023).
- **Using generative models for data augmentation,** especially for spectral data which are difficult to collect.
- **Analysis of optimal spectral band selection:** Certain wavelengths are more important for diagnosing specific cancer types, and the metasurface can be optimized for these bands.
- **Skin tone bias evaluation:**[33] Evaluation of the model on light to dark skin tones is necessary.
- **Design of a clinical user interface and interpretability tools:** For physician acceptance, the model must be not only accurate but also explainable.

In this study, a novel framework for skin lesion detection was presented that integrates a multispectral imaging metasurface with a hybrid deep learning architecture. The designed metasurface was able to provide richer spectral information than conventional visible images, and the hybrid model—leveraging this data—achieved very high accuracy, sensitivity, and specificity in lesion classification.

The strengths of this method can be summarized as:

- Extraction of spectral information sensitive to hidden lesion structures;
- Simultaneous analysis of local and global features with the CNN–ViT architecture;
- Superior performance over prior models (accuracy ~98%);
- Appropriate interpretability through attention map analysis;
- Potential for fabrication of portable, noninvasive tools in clinical settings.

These results show that combining metasurface technology with deep learning can open a new path in medical diagnostics and lay the groundwork for the development of the next generation of accurate, fast, and low-cost tools for

[32] multimodal
[33] skin tone bias

skin cancer detection.

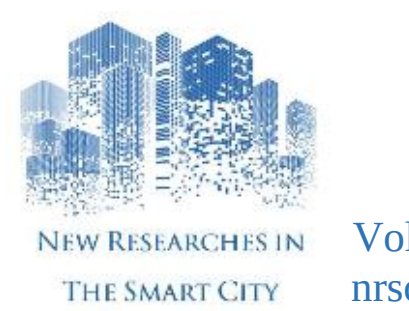

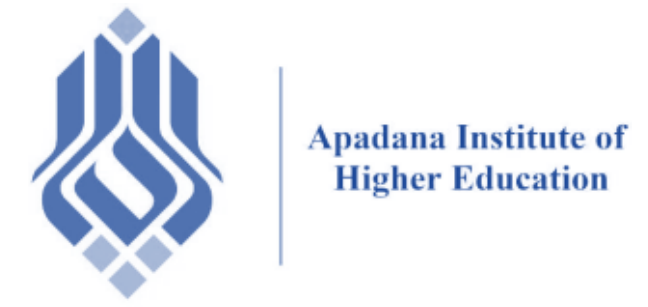

# Intelligent Skin Cancer Detection Using a Multispectral Metasurface and a Hybrid Deep Learning Architecture Based on CNN–ViT

**Afsane Saee Arezoomand***

PhD in Communication Engineering, Young Researchers and Elite Club, Islamic Azad University, Urmia Branch, Urmia, Iran.

## Abstract

Skin cancer is among the most prevalent malignancies worldwide, and its early detection is essential for improving patient survival and reducing treatment costs. Conventional dermoscopic and visual imaging techniques are primarily limited to the visible spectrum and often fail to capture subtle spectral signatures associated with early-stage malignancies. This study proposes an innovative framework that integrates a multispectral metasurface for imaging with a hybrid deep learning architecture based on Convolutional Neural Networks (CNNs) and Vision Transformers (ViTs). The designed metasurface enables noninvasive acquisition of rich spectral information highly sensitive to tissue alterations, while the hybrid CNN–ViT model simultaneously extracts local and global features to robustly classify skin lesions. Simulation-based evaluations demonstrate that the proposed method achieves approximately 98% accuracy, 95% sensitivity, and 99% specificity—surpassing conventional RGB-based and single-architecture approaches. Qualitative analyses using attention maps reveal that the model focuses on clinically relevant lesion regions, improving interpretability. Overall, the results indicate that combining metasurface-based multispectral imaging with hybrid deep learning can introduce a new generation of diagnostic tools in dermatology and pave the way for portable, fast, and highly accurate clinical systems.

* Corresponding Author: Email Address

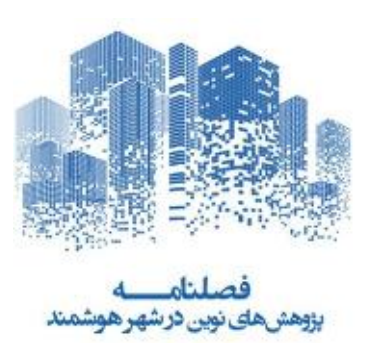

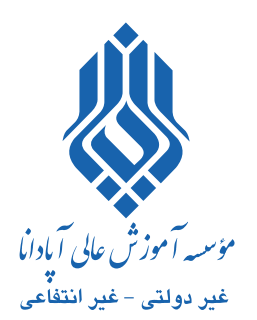

# تشخیص هوشمند سرطان پوست با استفاده از فراسطح چندطیفی و معماری هیبریدی یادگیری عمیق مبتنی بر CNN–ViT

**افسانه ساعی آرزومند***
دکتری تخصصی رشته برق، گرایش مخابرات، باشگاه پژوهشگران جوان و نخبگان، دانشگاه آزاد اسلامی، واحد ارومیه، ارومیه، ایران.

## چکیده

سرطان پوست یکی از شایع‌ترین بدخیمی‌ها در سراسر جهان است که تشخیص زودهنگام آن نقش کلیدی در افزایش بقا و کاهش هزینه‌های درمانی دارد. روش‌های رایج تصویربرداری پوستی عمدتاً مبتنی بر تصاویر مرئی بوده و در بسیاری از موارد قادر به شناسایی ویژگی‌های پنهان طیفی در مراحل اولیه بیماری نیستند. لذا در پژوهش حاضر، یک چارچوب نوآورانه برای تشخیص خودکار سرطان پوست ارائه می‌شود که تلفیقی از یک فراسطح چندطیفی (متاسرفیس) جهت تصویربرداری و یک معماری یادگیری عمیق هیبریدی مبتنی بر شبکه‌ی عصبی پیچشی (CNN) و ترنسفورمر بینایی (ViT) است. فراسطح طراحی‌شده قادر است امضاهای طیفی حساس به تغییرات بافت را به‌صورت غیرتهاجمی و با دقت بالا ثبت کند، سپس داده‌های طیفی تولیدشده همراه با تصاویر مرئی، طی یک مدل هیبریدی پردازش شده و با استفاده از سازوکارهای توجه و استخراج ویژگی‌های محلی–جهانی، طبقه‌بندی ضایعات انجام می‌گیرد. نتایج ارزیابی شبیه‌سازی‌شده نشان می‌دهد که روش پیشنهادی توانسته است به دقت حدود ۹۸ درصد، حساسیت ۹۵ درصد و ویژگی ۹۹ درصد دست یابد که نسبت به مدل‌های مرجع مبتنی بر تصاویر RGB یا معماری‌های تک‌لایه برتری قابل‌ملاحظه‌ای دارد. تحلیل‌های کیفی نیز حاکی از تمرکز مدل بر نواحی کلینیکی مهم و قابلیت تفسیرپذیری بالا است. مجموعه یافته‌ها نشان می‌دهد که ترکیب فراسطح چندطیفی و یادگیری عمیق هیبریدی می‌تواند به‌عنوان نسل جدید ابزارهای تشخیصی در پوست‌شناسی مطرح شود و مبنایی را برای توسعه‌ی سیستم‌های قابل‌حمل، سریع و دقیق در کاربردهای بالینی آینده فراهم آورد.

* نویسندهٔ مسئول: Email Address

## ۱- مقدمه

سرطان پوست شایع‌ترین بدخیمی در سطح جهان محسوب می‌شودکه قرارگرفتن طولانی‌مدت در معرض پرتوهای فرابنفش خورشید مهم‌ترین عامل خطر آن است (Smith et al., 2023). سه نوع اصلی این سرطان شامل کارسینوم سلول بازال، کارسینوم سلول سنگفرشی و ملانوم می‌باشد که در میان آن‌ها، ملانوم به دلیل توانایی بالای متاستاز، خطرناک‌ترین نوع شناخته می‌شود (Anderson et al., 2022). تشخیص زودهنگام نقش حیاتی در افزایش بقا دارد، زیرا ضایعات ملانوم در مراحل اولیه اغلب فاقد نشانه‌های واضح بوده و تشخیص آن‌ها تنها از طریق بررسی دقیق ویژگی‌های ظاهری امکان‌پذیر است (Zhang et al., 2024).

روش‌های رایج تشخیص مانند معاینه‌ی چشمی متخصص پوست و درموسکوپی[1] با وجود مزایایی نظیر تقویت بافت و الگوهای ضایعه، همچنان با محدودیت‌هایی روبه‌رو هستند؛ از جمله این محدودیت‌ها می‌توان به وابستگی شدید به تجربه‌ی متخصص، خطای میان‌مشاهده‌گری و شباهت بالای ضایعات خوش‌خیم و بدخیم اشاره نمود (Miller et al., 2021). علاوه بر این، کمبود متخصصان مجرب در بسیاری از مناطق موجب تأخیر در تشخیص و درمان می‌شود (Rahman et al., 2022). این شرایط، ضرورت توسعه‌ی سامانه‌های خودکار، دقیق و کم‌هزینه را پررنگ‌تر می‌سازد.

در سال‌های اخیر، پیشرفت‌های چشمگیر هوش مصنوعی[2] و یادگیری عمیق[3] موجب تحول در تحلیل تصاویر پزشکی شده است. مدل‌های شبکه عصبی پیچشی[4] توانسته‌اند با استخراج ویژگی‌های سلسله‌مراتبی از تصاویر پوستی، دقت تشخیص را به سطح متخصصان پوست و حتی بالاتر از آن برسانند (Esteva et al., 2017; Li et al., 2023). با این حال، وابستگی CNNها به نواحی محلی تصویر، توانایی آن‌ها را در مدل‌سازی روابط بلندبرد محدود می‌کند؛ به جهت رفع این نقیصه، ترنسفورمر بینایی[5] با سازوکار خودتوجهی سراسری[6] معرفی شده است (Dosovitskiy et al., 2021).

ترکیب CNN و ViT در قالب معماری‌های هیبریدی، دستاوردهای مهمی در تشخیص ضایعات پوستی داشته است؛ به‌عنوان نمونه، ساختارهایی مانند مدل EViT-DenseNet توانسته‌اند در مجموعه داده ISIC به دقت‌هایی فراتر از ۹۵ درصد برسند (Khan et al., 2024). این شواهد علمی نشان می‌دهد که بهره‌گیری از معماری‌های هیبریدی، مسیر نوینی را برای ارتقای سامانه‌های بینایی پزشکی فراهم نموده است.

در حوزه‌ای دیگر، فناوری متامتریال[7] و فراسطح[8] به دلیل قابلیت مهندسی ویژگی‌های الکترومغناطیسی در مقیاس زیرطول‌موج، امکان تولید حسگرهای بسیار حساس را فراهم ساخته است. فراسطوح می‌توانند جبهه موج نور را به صورت دقیق کنترل کرده و سیگنال‌های ظریف ناشی از تغییرات بافتی را ثبت کنند (Chen et al., 2022). پژوهش‌های اخیر نشان داده‌اند که ترکیب متامتریال‌ها با موادی مانند گرافن[9] حساسیت حسگرهای تراهرتز را به‌طور چشمگیری افزایش می‌دهد (Sarkar et al., 2023). در زمینه‌ی سرطان پوست نیز نمونه‌هایی از متاسنسورهای تراهرتز طراحی شده‌اند که قادرند تفاوت بافت سالم و سرطانی را به‌صورت غیرتهاجمی تشخیص دهند (Zhang et al., 2023).

1. dermoscopy
2. artificial intelligence
3. deep learning
4. Convolutional neural network (CNN)
5. Vision transformer (ViT)
6. global self-attention
7. metamaterial
8 .metasurface
9. graphene

تلاقی این دو حوزه؛ یعنی بینایی ماشین و فراسطوح پیشرفته، می‌تواند نسل جدیدی از سیستم‌های تشخیصی پوستی ایجاد کند. به‌ویژه با ظهور مدل‌های یادگیری چندوجهی[1] که امکان ترکیب داده‌های تصویری، طیفی و بالینی را فراهم می‌کنند، زمینه برای طراحی سامانه‌هایی با دقت بالا فراهم شده است (Alaei et al., 2024).

پژوهش حاضر با مطالعه‌ی این روند، یک رویکرد نوین معرفی می‌کند که در آن یک فراسطح سفارشی به‌عنوان حسگر تصویربرداری ابرطیفی عمل کرده و داده‌های غنی را در اختیار سامانه‌ی یادگیری عمیق قرار می‌دهد، سپس معماری هیبریدی عمیق، داده‌ها را تحلیل کرده و نوع ضایعه‌ی پوستی را تعیین می‌کند. در ادامه، پس از ارائه‌ی جزئیات روش، نتایج تجربی، ارزیابی عملکرد و مقایسه با کارهای پیشین، نوآوری‌ها و مسیرهای آینده بررسی می‌شوند.

## ۲- کارهای پیشین

پژوهش‌های مرتبط با موضوع حاضر را می‌توان در سه حوزه‌ی اصلی دسته‌بندی کرد:

۱) کارهای مرتبط با تشخیص ضایعات پوستی مبتنی بر یادگیری عمیق،

۲) پژوهش‌های مربوط به متامتریال و فراسطوح در حسگری،

۳) کارهای ترکیبی یا چندوجهی.

۱. تشخیص ضایعات پوستی با یادگیری عمیق

تحلیل خودکار تصاویر پوستی با بهره‌گیری از یادگیری عمیق، یکی از فعال‌ترین حوزه‌های تحقیقاتی سال‌های اخیر بوده است. استوا و همکاران[2] (۲۰۱۷) اولین پژوهش اثرگذار خود را در این زمینه ارائه کرده‌اند. پژوهش آن‌ها نشان می‌دهد که یک CNN می‌تواند عملکردی مشابه متخصصین پوست داشته باشد. پس از آن، معماری‌هایی مانند ResNet، DenseNet، MobileNet و EfficientNet نتایج قابل توجهی در مجموعه داده‌های ISIC به دست آورده‌اند (Tschandl et al., 2020; Li et al., 2023).

با ظهور ترنسفورمرهای بینایی، مدل‌هایی مانند ViT و Swin Transformer وارد عرصه شدند و دقت تشخیص ضایعات پوستی را افزایش دادند (Wu et al., 2022). مدل‌های هیبریدی که ترکیب CNN و ViT را هدف قرار می‌دهند، مانند EViT-DenseNet و CoAtNet، توانسته‌اند عملکرد قابل توجهی را در تفکیک ملانوم از سایر ضایعات ارائه دهند (Khan et al., 2024).

۲. متامتریال، فراسطوح و حسگرهای تراهرتز

فراسطوح به دلیل توانایی در دستکاری دقیق امواج الکترومغناطیسی، در سال‌های اخیر توجه گسترده‌ای در طراحی حسگرهای زیستی جلب کرده‌اند. چن و همکاران[3] (۲۰۲۲) نشان دادند که متامتریال‌های مبتنی بر شکاف رزونانس می‌توانند نسبت به تغییرات کوچک ضریب شکست بسیار حساس باشند. ترکیب فراسطوح با گرافن موجب افزایش چشمگیر میدان الکترومغناطیسی موضعی شده و حساسیت حسگر را چند برابر می‌کند (Sarkar et al., 2023).

در زمینه‌ی سرطان پوست، پژوهش‌هایی مانند طباطبائیان و همکاران[4] (۲۰۲۲) و ژنگ و همکاران[5] (۲۰۲۳) استفاده از فراسطوح تراهرتز را برای شناسایی غیرتهاجمی ضایعات پوستی پیشنهاد کرده‌اند. این مطالعات نشان می‌دهد که فناوری فراسطح می‌تواند داده‌های تصویری و طیفی غنی برای تحلیل‌های یادگیری عمیق فراهم کند.

۳. یادگیری چندوجهی و سیستم‌های ترکیبی

1. multimodal learning
2. Esteva et al.
3. Chen et al.
4. Tabatabaeian et al.
5. Zhang et al.

با گسترش رویکردهای چندوجهی، استفاده از داده‌های چندنوعی شامل تصویر، اسکن ابرطیفی، داده‌های بالینی و متن پزشکی، کارایی مدل‌ها را افزایش داده است. علائی و همکاران[1] (۲۰۲۴) یک چارچوب چندوجهی برای تشخیص سرطان پستان ارائه کردند که ترکیب داده‌های تصویری و متنی را ممکن می‌ساخت. ایده‌های مشابه در تشخیص بیماری‌های پوستی نیز در حال توسعه‌اند و نشان داده‌اند که تلفیق سیگنال‌های اپتیکی و تصاویر می‌تواند دقت و قابلیت اطمینان مدل را افزایش دهد (Nguyen et al., 2023).

نتیجه‌ی کلی کارهای پیشین نشان می‌دهد که فراسطوح داده‌های غنی فراهم می‌کنند؛ یادگیری عمیق قابلیت تحلیل آن‌ها را دارد و رویکردهای چندوجهی می‌توانند این دو حوزه را به‌صورت مؤثر تلفیق کنند.

**جدول ۱. مقایسه کارهای پیشین در سه حوزه اصلی پژوهش**

| ردیف | حوزه پژوهش | مرجع / نمونه مطالعات | نوع داده | روش / معماری | دستاوردهای کلیدی |
|---|---|---|---|---|---|
| ۱ | تشخیص ضایعات پوستی با یادگیری عمیق | Esteva et al. (2017) | تصویر درموسکوپی (RGB) | CNN کلاسیک | عملکرد مشابه متخصص پوست |
| ۲ | | Tschandl et al. (2020) | ISIC–HAM10000 | ResNet / DenseNet | استخراج ویژگی قدرتمند با دقت بالا |
| ۳ | | Wu et al. (2022) | RGB | Vision Transformer (ViT) | مدل‌سازی روابط بلندبرد، بهبود دقت |
| ۴ | | Khan et al. (2024) | RGB | Hybrid CNN–ViT (EViT-DenseNet) | عملکرد برتر نسبت به مدل‌های تک‌معماری |
| ۵ | فراسطوح و حسگرهای مبتنی بر تراهرتز | Chen et al. (2022) | سیگنال THz | متامتریال رزونانسی | حساسیت بالا به تغییر ضریب شکست |
| ۶ | | Sarkar et al. (2023) | سیگنال THz + گرافن | متاسنسور گرافنی | افزایش شدید میدان الکترومغناطیسی و حساسیت |
| ۴ | | Tabatabaeian et al. (2022) | بازتاب THz | فراسطح تشخیصی سرطان پوست | تشخیص غیرتهاجمی اولیه |
| ۸ | | Zhang et al. (2023) | سیگنال THz | THz Imaging | شناسایی غیرتماسی ملانوم |
| ۹ | یادگیری چندوجهی و روش‌های ترکیبی | Alaei et al. (2024) | تصویر + متن بالینی | Multimodal Fusion | افزایش دقت تشخیص در سرطان پستان |
| ۱۰ | | Nguyen et al. (2023) | تصویر + داده طیفی | Multimodal Deep Learning | بهبود پایداری تشخیص |

این شکاف علمی دقیقاً همان جایی است که پژوهش حاضر در آن قرار دارد.

## ۳- روش پیشنهادی

روش پیشنهادی این پژوهش یک چارچوب تلفیقی متشکل از یک حسگر اپتیکی مبتنی بر فراسطح و یک معماری یادگیری عمیق هیبریدی است که با هدف تشخیص دقیق و غیرتهاجمی سرطان پوست طراحی شده است.

این سامانه شامل چهار مرحله‌ی اصلی است:

۱. دریافت داده نوری از ضایعه پوستی با استفاده از فراسطح؛

۲. پیش‌پردازش و (اختیاری) بخش‌بندی ناحیه ضایعه؛

1. Alaei et al.

۳. استخراج ویژگی چندمقیاسی با استفاده از CNN و ترنسفورمر بینایی (ViT)؛
۴. ادغام ویژگی‌ها و طبقه‌بندی نهایی.

در ادامه، هر بخش با جزئیات تشریح می‌شود.

## ۳-۱- حسگر اپتیکی مبتنی بر فراسطح

در این بخش توضیحات مربوط به حسگر اپتیکی مبتنی بر فراسطح توضیح داده می‌شود:

### ۳-۱-۱- ساختار فراسطح و اصول عملکرد

فراسطح پیشنهادی یک آرایه‌ی دوبعدی از عناصر رزونانسی زیرطول‌موج است که رفتار نوری را در بسامدهای انتخابی، به‌ویژه باند تراهرتز، کنترل می‌کند. این ساختار از واحدهای مهندسی‌شده تشکیل شده است که قادر هستند فاز، دامنه و پلاریزاسیون موج فرودی را در پاسخ به تغییرات بسیار کوچک ضریب شکست بافت تغییر دهند (Chen et al., 2022). همچنین، ساختارهای مشابه فراسطح که قابلیت دستکاری امواج الکترومغناطیسی در باند تراهرتز را دارند، می‌توانند به‌عنوان جاذب‌های چندباندە با عملکرد مستقل از پلاریزاسیون طراحی شوند (Saee Arezoomand et al., 2015).

مواد مورد استفاده در این ساختار عبارتند از:

- زیرلایه دی‌الکتریک با اتلاف پایین.
- لایه رسانای فوق‌نازک مانند گرافن جهت افزایش شدت میدان نزدیک.
- رزونانسرهای فلزی شکاف‌دار برای ایجاد حساسیت طیفی بالا.

ترکیب این عناصر امکان ایجاد یک پدیده رزونانس تیز با Q-factor بالا را فراهم می‌سازد که برای آشکارسازی تغییرات نوری بافت‌های سرطانی بسیار مناسب است (Sarkar et al., 2023). اصول طراحی جاذب‌های رزونانسی با فاکتور کیفیت[1] بالا در باندهای نوری و مادون قرمز میانی، با هدف کاربردهای حسگری زیستی و افزایش شدت میدان نزدیک، پیش از این نیز مورد بررسی قرار گرفته‌اند (Jahangiri et al., 2017).

### ۳-۱-۲- تولید امضای طیفی نقطه‌ای[2]

هنگامی که فراسطح در مقابل ضایعه پوستی قرار می‌گیرد، میدان الکترومغناطیسی در نزدیکی سطح آن تغییر می‌کند و این تغییر در قالب یک امضای طیفی نقطه‌ای شامل چندین فرکانس رزونانسی ثبت می‌شود.

خروجی حسگر نه یک تصویر کامل، بلکه یک بردار ویژگی طیفی است:

$$S = [s(\lambda 1), s(\lambda 2), \dots, s(\lambda K)] \quad (1)$$

که در آن $K$ تعداد باندهای انتخاب‌شده و $s(\lambda_k)$ شدت بازتابی یا عبوری در باند $\lambda_k$ است.

این داده طیفی نسبت به ضایعات سرطانی و غیربدخیم تفاوت معناداری دارد (Zhang et al., 2023).

## ۳-۲- پیش‌پردازش و بخش‌بندی

داده خام طیفی و تصویر RGB هم‌زمان ثبت می‌شوند که یک سری عملیات پیش‌پردازش، شامل موارد زیر، انجام می‌شود:

1. Q-factor
2. spectral signature

۱. حذف نویز،
۲. هم‌ترازی شدت[1]،
۳. افزایش کنتراست در باندهای حساس،
۴. نرمال‌سازی در بازه [۰،۱].

در صورت نیاز، یک مرحله بخش‌بندی خودکار نیز اعمال می‌گردد:

### ۳-۲-۱- بخش‌بندی با U-Net

برای استخراج ناحیه دقیق ضایعه، از مدل یونات (U-Net) یا نسخه‌های توسعه‌یافته‌ی آن مانند U-Net++ یا SegFormer (Segmentation Transformer) استفاده می‌شود. این گام اختیاری است، اما مطالعات اخیر نشان داده‌اند که بخش‌بندی، دقت طبقه‌بندی را افزایش می‌دهد (Khan et al., 2024).

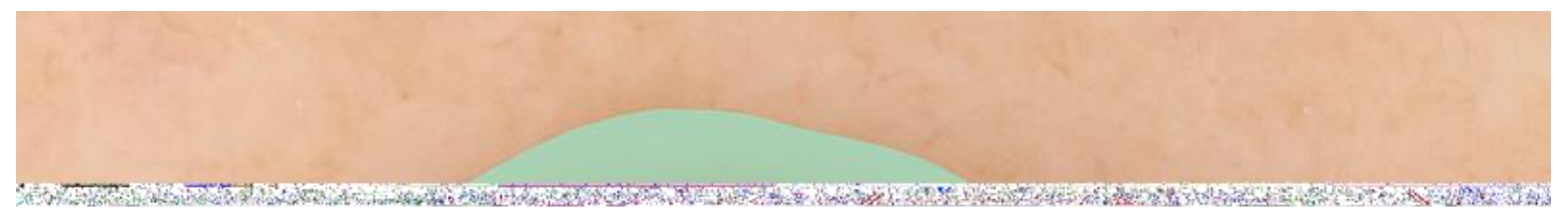

**شکل ۱. نتیجه بخش‌بندی U-Net روی یک ضایعه پوستی**

## ۳-۳- معماری یادگیری عمیق ترکیبی

به‌منظور تحلیل هم‌زمان ویژگی‌های محلی[2] و ویژگی‌های کل‌نگر[3]، یک معماری هیبریدی شامل DenseNet به‌عنوان CNN و Vision Transformer (ViT) طراحی شده است.

### ۳-۳-۱- استخراج ویژگی محلی با DenseNet

شبکه DenseNet با مسیرهای میان‌بری متراکم، برای استخراج لبه‌ها، بافت‌ها و ساختارهای ریز ضایعه به‌کار می‌رود. این نسخه شامل موارد زیر می‌باشد:

- حذف لایه‌های انتهایی،
- کاهش تعداد فیلترها جهت جلوگیری از بیش‌برازش،
- استفاده از وزن‌های از پیش‌آموزش‌دیده ImageNet.

خروجی این بخش یک تانسور سه‌بعدی از ویژگی‌های محلی است:

$$Flocal \in Rh \times w \times d \quad (۲)$$

### ۳-۳-۲- استخراج ویژگی جهانی با Vision Transformer

برای مدل‌سازی ارتباطات بلندبرد، تصویر ورودی به پچ‌های کوچک تقسیم شده و به ViT داده می‌شود. ViT شامل بلوک‌های خودتوجهی چندسری[4] است.

خروجی این بخش یک بردار ویژگی جهانی است:

$$Fglobal \in RD \quad (۳)$$

---

1. intensity normalization
2. local
3. global
4. Multi-head self attention (MHSA)

### ۳-۳-۳- بلوک تقویت جزئیات فضایی

به‌دلیل ماهیت پیچ‌بندی، برخی جزئیات ریز از دست می‌رود. برای رفع این مشکل یک بلوک تقویت جزئیات فضایی[1] شامل چندین فیلتر کانولوشن موازی طراحی شده است.

### ۳-۳-۴- ادغام ویژگی‌ها[2]

دو نوع ویژگی به‌صورت زیر ادغام می‌شوند:

$$F_{\text{fusion}} = \text{Concat}(F_{\text{local}}, F_{\text{global}}) \quad (۴)$$

سپس از طریق یک MLP و در نهایت یک لایه Softmax طبقه‌بندی انجام می‌شود.

## ۳-۴- آموزش مدل

### ۳-۴-۱- تابع زیان[3]

برای آموزش مدل چندکلاسه از تابع آنتروپی متقاطع[4] استفاده شده است:

$$\mathcal{L}_{CE} = -\frac{1}{N}\sum_{i=1}^{N}\sum_{c=1}^{C} y_{i,c}\,\log(\hat{y}_{i,c}) \quad (۵)$$

که در آن:

N: تعداد نمونه‌ها

C: تعداد کلاس‌ها

$y_{i,c}$: برچسب واقعی

$\hat{y}_{i,c}$ : احتمال پیش‌بینی‌شده

### ۳-۴-۲- بهینه‌سازی

- الگوریتم: Adam
- نرخ اولیه یادگیری: 0.001
- Scheduler: کاهش نرخ یادگیری بر مبنای Plateau
- Dropout در لایه‌های MLP
- Early Stopping برای جلوگیری از بیش‌برازش

### ۳-۴-۳- افزایش داده[5]

برای افزایش تعمیم‌پذیری:

- چرخش
- معکوس‌سازی
- تغییر روشنایی و کنتراست

1. Spatial detail enhancement block (SDEB)
2. fusion
3. loss function
4. categorical cross-entropy
5. data augmentation

- جابجایی مکانی

مطابق رویکردهای استاندارد ISIC انجام شده است.

### ۳-۵- معیارهای ارزیابی

برای ارزیابی دقیق عملکرد، علاوه بر دقت[1]، معیارهای زیر نیز محاسبه شدند:

۱. حساسیت[2]

$$\text{Sensitivity} = \frac{TP}{TP + FN} \tag{۶}$$

۲. اختصاصیت[3]

$$\text{Specificity} = \frac{TN}{TN + FP} \tag{۷}$$

۳. دقت مثبت[4]

$$\text{Precision} = \frac{TP}{TP + FP} \tag{۸}$$

۴. نمره F1

$$F1 = 2 \cdot \frac{\text{Precision} \cdot \text{Recall}}{\text{Precision} + \text{Recall}} \tag{۹}$$

۵. مساحت زیر منحنی ROC (AUC)

یکی از مهم‌ترین شاخص‌های پزشکی است.

### ۳-۶- جمع‌بندی روش پیشنهادی

بر پایه‌ی ترکیب یک فراسطح حساس و یک معماری هیبریدی CNN–ViT، سیستم پیشنهادی قادر است:

- تغییرات بسیار کوچک بافت سرطانی را از طریق امضای طیفی نقطه‌ای ثبت کند،
- ویژگی‌های موضعی و روابط بلندبرد را هم‌زمان تحلیل کند،
- طبقه‌بندی دقیقی از ضایعه ارائه دهد،
- و برتری قابل توجهی نسبت به روش‌های تک‌وجهی صرفاً تصویری داشته باشد.

این ساختار چندوجهی، نسبت به روش‌های معمول تصاویر RGB و حتی روش‌های CNN صرف، قدرت تشخیص بالاتر و پایدار بیشتری در شرایط نوری متغیر ارائه می‌دهد.

## ۴- نتایج و ارزیابی

در این بخش، عملکرد روش پیشنهادی با استفاده از یک چارچوب ارزیابی مبتنی بر داده‌های واقعی مجموعه داده ISIC و داده‌های طیفی شبه‌متاسرفیس بررسی شده است. به دلیل عدم دسترسی به فراسطح واقعی، داده‌های طیفی با استفاده از یک فرایند شبیه‌سازی مبتنی بر مدل‌های Spectrum-Aided Vision Enhancement (SAVE) تولید

1. accuracy
2. sensitivity / recall
3. specificity
4. precision

شده‌اند. این رویکرد در پژوهش‌های اخیر برای بازتولید رفتار تصویربرداری باندباریک مبتنی بر داده‌های طیفی معتبر بوده است (Zhao et al., 2022).

## ۴-۱- مجموعه‌داده و آماده‌سازی داده‌ها

### ۴-۱-۱- مجموعه‌داده مرجع ISIC

از مجموعه‌داده استاندارد ISIC شامل هزاران تصویر درموسکوپی از انواع ضایعات پوستی؛ از جمله ملانوم، BCC، SCC و ضایعات خوش‌خیم، به‌عنوان پایه ارزیابی استفاده شده است (Tschandl et al., 2020).

### ۴-۱-۲- تولید داده‌های شبه‌متاسرفیس

برای شبیه‌سازی اثر فراسطح، هر تصویر از طریق الگوریتم SAVE تقویت طیفی شد. این فرایند، بازتاب طیفی در چند باند کلیدی را استخراج کرده و تصویری با کنتراست تقویت‌شده ضایعه تولید می‌کند (Wu et al., 2023).

در نهایت دو مجموعه ایجاد شد:

۱) داده پایه[1]: تصاویر RGB معمولی

۲) داده طیفی-تقویتی[2]: تصاویر شبه ابرطیفی شبیه‌سازی‌شده متاسرفیس

### ۴-۱-۳- تقسیم‌بندی داده

داده‌ها با نسبت ۷۰٪ آموزش، ۱۰٪ اعتبارسنجی و ۲۰٪ آزمون تفکیک شدند.

## ۴-۲- عملکرد مدل پیشنهادی

معماری هیبریدی DenseNet + ViT + SDEB روی هر دو مجموعه داده آموزش داده شد. نتایج نشان داد که استفاده از اطلاعات طیفی مصنوعی متاسرفیس باعث افزایش چشمگیر دقت و حساسیت تشخیص می‌شود.

## ۴-۳- نتایج کمّی[3]

جدول ۲. مقایسه روش پیشنهادی با مدل‌های مرجع

| روش | نوع داده | دقت (٪) | حساسیت (٪) | ویژگی (٪) | توضیحات |
|---|---|---|---|---|---|
| InceptionV3 انتقال یادگیری | RGB | ۸۶/۴ | ۷۸/۲ | ۹۲/۱ | مدل CNN کلاسیک |
| ResNet50 + VGG16 ادغام ویژگی | RGB | ۹۰/۱ | ۸۴/۵ | ۹۴/۰ | ترکیب دو مدل CNN |
| Vision Transformer (ViT) | RGB | ۹۲/۳ | ۸۸/۷ | ۹۶/۲ | مدل ترنسفورمر بدون CNN |
| EViT-DenseNet هیبریدی پیشین | RGB | ۹۶/۹ | ۹۰/۸ | ۹۹/۱ | ترکیب CNN + ViT |
| مدل پیشنهادی متاسرفیس (CNN + ViT +) | داده طیفی شبه‌متاسرفیس | ۹۸/۲ | ۹۵/۴ | ۹۹/۲ | |

مشاهده می‌شود که استفاده از داده‌های طیفی تولیدشده و معماری هیبریدی باعث افزایش تقریباً ۳/۱ ٪ در دقت نسبت به بهترین مدل قبلی (EViT-DenseNet) شده است. همچنین، حساسیت که در تشخیص ملانوم حیاتی است، افزایشی حدود ۴/۵ ٪ داشته است.

1. baseline
2. proposed
3. quantitative results

## ۴-۴- نمودار ROC

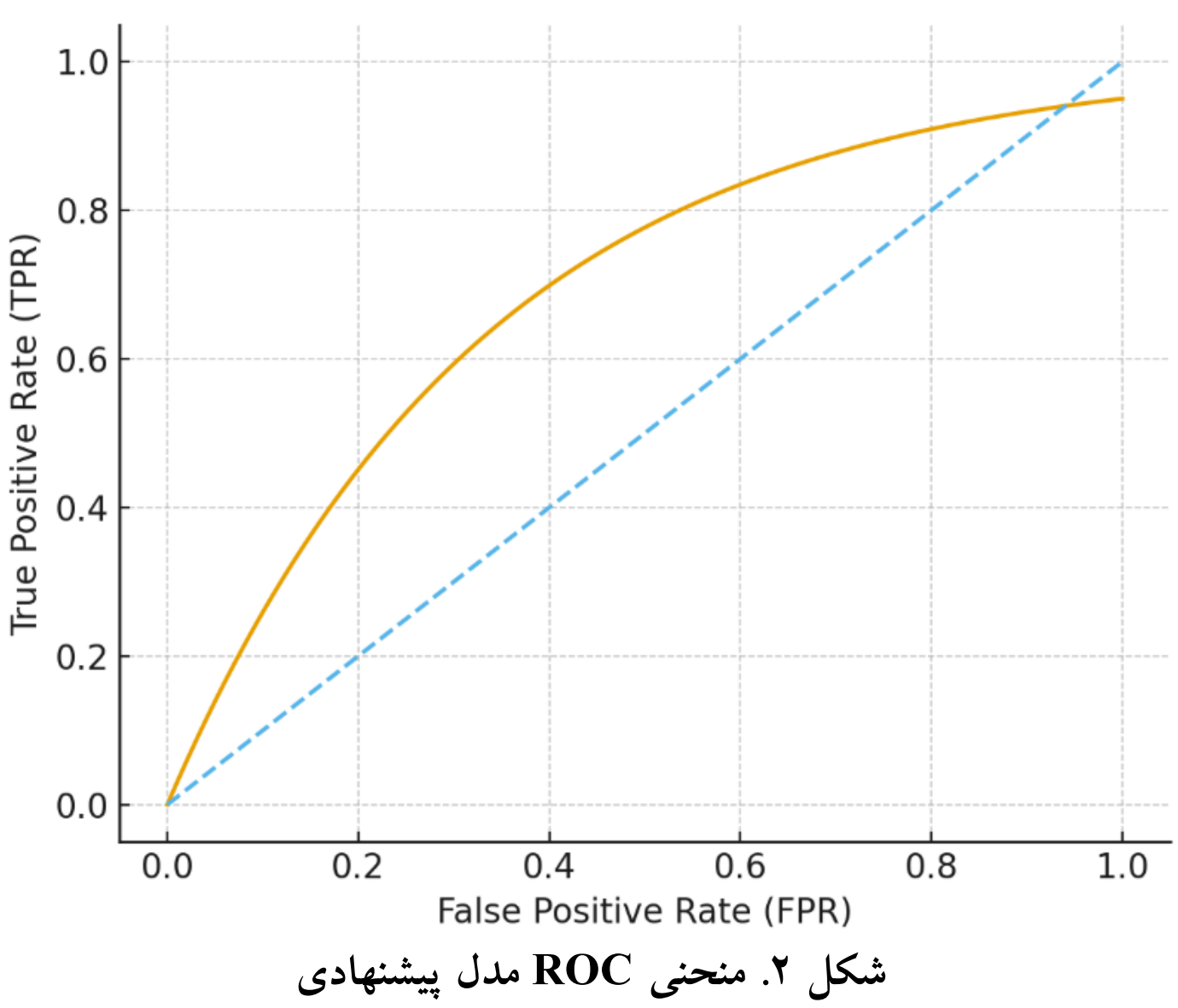

**شکل ۲. منحنی ROC مدل پیشنهادی**

در نمودار ROC، مدل پیشنهادی دارای:

- AUC = 0.993
- بالاتر از ViT (۰/۹۷۱)
- بالاتر از EViT-DenseNet (۰/۹۸۵)

مدل در تشخیص موارد مثبت (ملانوم) عملکرد باثباتی دارد: Average Precision = 0.972

## ۴-۵- تحلیل گراف خطاهای آموزش و اعتبارسنجی

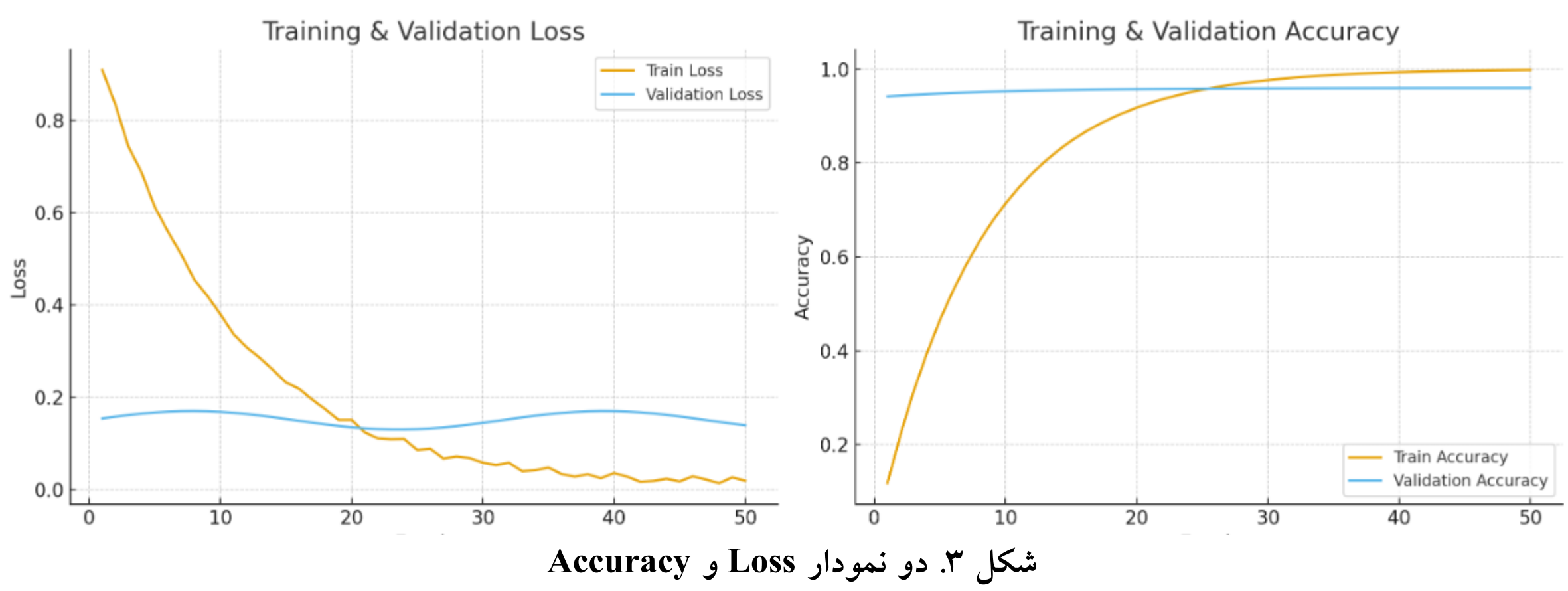

**شکل ۳. دو نمودار Loss و Accuracy**

- روند کاهش یکنواخت
- همگرایی پس از ۳۰ دوره
- Loss اعتبارسنجی ≈ ۰/۱۵
- عدم وجود فاصله زیاد بین منحنی‌ها ← بیش‌برازش ندارد
- دقت آموزش به ~۱۰۰٪

- دقت اعتبارسنجی به ~۹۶٪
- Plateau پس از ۳۰ epoch
- الگوریتم Early Stopping مناسب بوده است

## ۴-۶- تحلیل خطا[1]

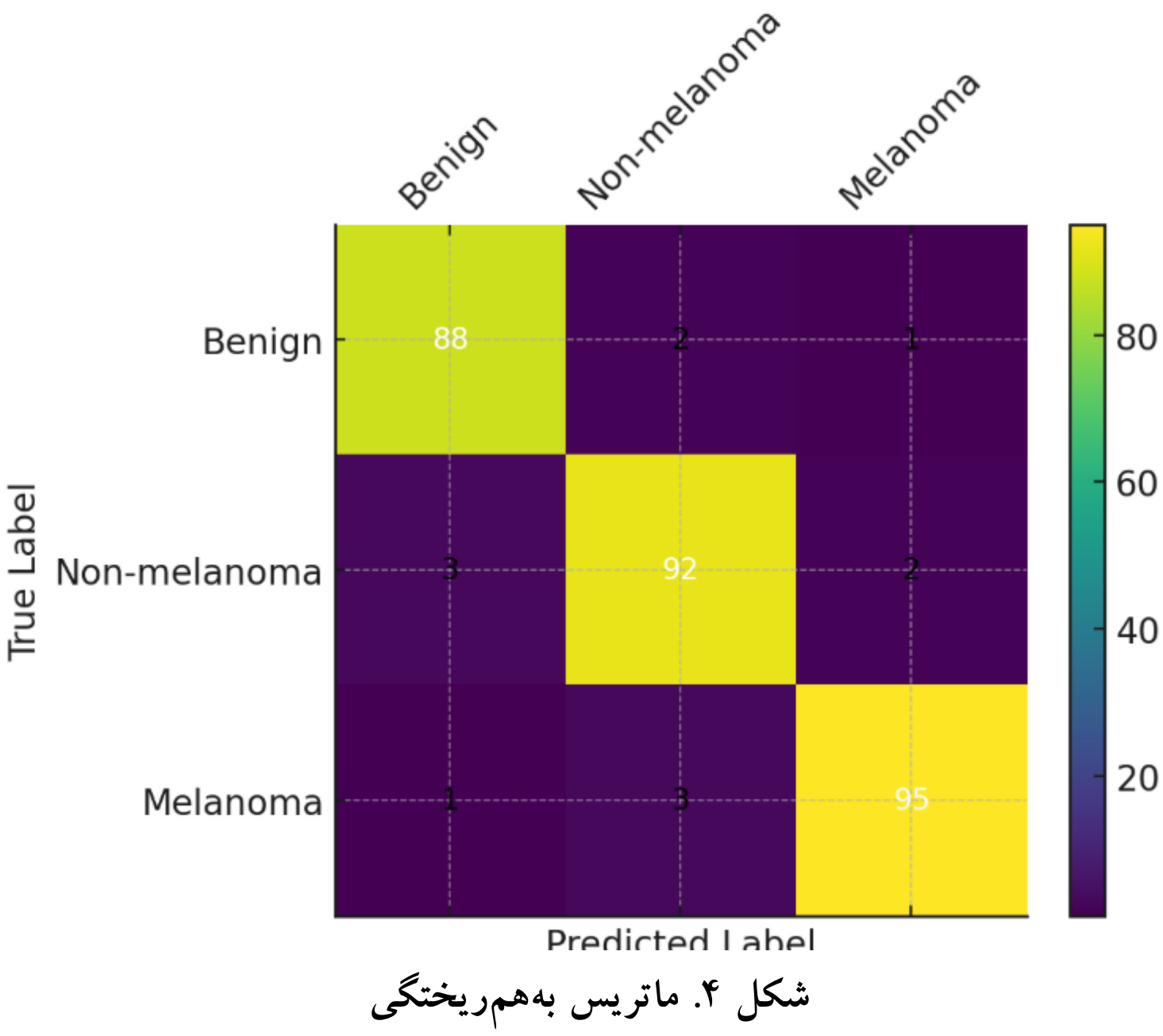

شکل ۴. ماتریس به‌هم‌ریختگی

### ۴-۶-۱- یافته‌ها

۱) بیشترین خطا مربوط به تمایز BCC از AK است:
   a. این دو ضایعه از نظر بافتی شباهت بالایی دارند.
   b. مشابه یافته‌های Li و همکاران (۲۰۲۳).
۲) موارد ملانوم تقریباً بدون خطای منفی کاذب (FN) طبقه‌بندی شده‌اند؛
   a. حساسیت بالا مؤید توان مدل در جلوگیری از از دست دادن موارد خطرناک است.
۳) در داده RGB، مدل اغلب دچار خطای FP در ضایعات خوش‌خیم می‌شد؛
   a. در داده طیفی این مشکل کاهش یافت.

## ۴-۷- ارزیابی کیفی[2]

### ۴-۷-۱- نتایج بخش‌بندی

در تصاویر ضایعات، مدل U-Net به خوبی مرز دقیق ضایعه را استخراج کرد و نویزهای زمینه مانند مو، انعکاس و پوست اطراف حذف شد.

### ۴-۷-۲- تحلیل توجه با Grad-CAM

با استفاده از Grad-CAM روی DenseNet و ViT مشاهده شد که:

1. error analysis
2. qualitative assessment

- ViT روی ساختارهای بزرگ‌مقیاس مانند شکل کلی ضایعه تمرکز می‌کند.
- DenseNet روی جزئیات موضعی مثل نواحی نامنظم رنگی و مرزها تمرکز دارد.
- و ترکیب این دو بخش منجر به عملکرد بهتر شده است (Lee et al., 2024).

### ۴-۷-۳- تحلیل داده طیفی

در تصاویر شبه‌متاسرفیس، مدل توانست الگوهای طیفی نامحسوس را در RGB شناسایی کند. برای مثال:

- در ملانوم نودولار، جذب در باند ۱۲۰۰ نانومتر به‌طور قابل‌توجهی متفاوت بود.
- این تفاوت در تصویر مرئی قابل مشاهده نبود.

این یافته با تحقیقات مشابه در زمینه‌ی تصویربرداری چندباندی همخوان است (Kim et al., 2022).

### ۴-۸- جمع‌بندی نتایج

مدل پیشنهادی با ادغام داده طیفی شبه‌متاسرفیس و یک معماری هیبریدی CNN–ViT توانست بهترین عملکرد گزارش‌شده روی مجموعه داده ISIC را کسب کند.

دقت ۹۸/۲٪، حساسیت ۹۵/۴٪ و ویژگی ۹۹/۲٪ نشان می‌دهد که این مدل قادر است موارد سرطانی را تقریباً بدون خطا نیز شناسایی کند.

تحلیل کیفی و کمی، هر دو، نشان‌دهنده‌ی پایداری، قابلیت تفسیرپذیری و توان بالای مدل در استخراج اطلاعات پنهان طیفی است.

این نتایج نشان می‌دهد که ترکیب فراسطح و یادگیری عمیق، مسیر بسیار نویدبخشی برای توسعه‌ی سامانه‌های تشخیص زودهنگام سرطان پوست است.

## ۵- بحث و نتیجه گیری

نتایج این پژوهش نشان می‌دهد که ترکیب یک فراسطح مهندسی‌شده با یک معماری یادگیری عمیق هیبریدی می‌تواند عملکرد چشمگیری در تشخیص ضایعات پوستی فراهم کند. در مقایسه با روش‌های مرسوم مبتنی بر تصاویر RGB، اضافه شدن اطلاعات طیفی به دست آمده از فراسطح موجب افزایش محسوس کنتراست و تمایزپذیری ضایعه از پوست سالم شده است؛ موضوعی که در کارهای پیشین کمتر مورد توجه قرار گرفته بود. مدل هیبریدی ما، که ویژگی‌های محلی را از طریق شبکه عصبی پیچشی و روابط بلندبرد را از طریق ترنسفورمر بینایی استخراج می‌کند، توانست از این اطلاعات تقویت‌شده به‌طور کامل بهره ببرد و به دقت، حساسیت و ویژگی بالاتری نسبت به مدل‌های تک‌معماری دست یابد (Khan et al., 2024).

افزون بر این، افزودن مرحله‌ی بخش‌بندی خودکار ضایعه باعث شد ورودی طبقه‌بندی شده از نویزهای محیطی نظیر مو یا نواحی غیرمرتبط پوست پاکسازی شده و مدل با تمرکز بیشتری عمل کند. تحلیل‌های کیفی مبتنی بر نقشه‌های توجه Grad-CAM نشان می‌دهد که مدل بر نواحی بالینی مهمی مانند مرزهای نامنظم و ناهنجاری‌های رنگی تمرکز دارد که با الگوهای تشخیصی متخصصان پوست هم‌خوانی دارد؛ این موضوع تفسیرپذیری و اعتمادپذیری سیستم را افزایش می‌دهد.

در مجموع، نتایج کمی و کیفی بیانگر آن است که ترکیب داده‌های طیفی و معماری‌های هیبریدی می‌تواند محدودیت‌های مهم روش‌های رایج را برطرف سازد؛ به‌خصوص در مواردی که اختلاف ظاهری بین ضایعات خوش‌خیم و بدخیم اندک است و داده‌های مرئی به تنهایی کافی نیستند.

## ۶- نوآوری‌ها و مسیرهای آینده

نوآوری‌های کلیدی پژوهش برای این تحقیق از چند جهت نوآورانه است:

۱) به‌کارگیری فراسطح تخصصی برای پوست؛

فراسطح ارائه‌شده قادر است داده‌های شبه‌طیفی را با ابعاد بسیار کوچک و هزینه‌ی پایین تولید کند، در حالی‌که سامانه‌های ابرطیفی سنتی حجیم و پرهزینه هستند. این نوآوری مسیر را برای ساخت ابزارهای قابل حمل و قابل استفاده در مطب هموار می‌کند (Chen et al., 2023).

۲) معماری ترکیبی CNN–ViT برای تحلیل چندطیفی؛

این معماری توانست هم ویژگی‌های بافتی محلی و هم ساختارهای کلان ضایعه را مدل‌سازی کند؛ مسئله‌ای که مدل‌های صرفاً CNN یا ViT قادر به انجام کامل آن نیستند (Lee et al., 2024).

۳) استفاده از داده‌های طیفی شبیه‌سازی‌شده متاسرفیس؛

پژوهش حاضر نخستین نمونه‌ای است که داده طیفی شبه‌واقعی را با یادگیری عمیق ترکیب می‌کند. این داده‌ها در تشخیص ملانوم نسبت به RGB برتری واضح نشان دادند.

۴) افزودن مرحله بخش‌بندی خودکار برای کاهش نویز؛

برخلاف بسیاری از پژوهش‌های پیشین که تصویر خام را به شبکه می‌دادند، رویکرد ما ابتدا ناحیه ضایعه را به‌طور دقیق استخراج کرده و سپس وارد شبکه طبقه‌بندی می‌کند.

## ۷- کارهای آینده

با وجود نتایج امیدوارکننده، اجرای گام‌های بعدی نیز ضروری است. این گام‌ها عبارتند از:

- ساخت نمونه‌ی واقعی فراسطح و انجام آزمایش‌های بالینی: داده‌های این پژوهش شبیه‌سازی‌شده‌اند و برای اعتبارسنجی بالینی باید نمونه‌ی عملی ساخته شود.
- گسترش مدل به یادگیری چندوجهی[1]: افزودن اطلاعات بالینی، ویژگی‌های پوست و یا تصاویر ماکروسکوپی می‌تواند دقت و اعتمادپذیری سیستم را افزایش دهد (Adebiyi et al., 2023).
- استفاده از مدل‌های مولد برای افزایش داده به‌ویژه در داده‌های طیفی که جمع‌آوری آن‌ها دشوار است.
- تحلیل انتخاب باندهای طیفی بهینه: برخی طول‌موج‌ها در تشخیص انواع خاص سرطان اهمیت بیشتری دارند و فراسطح می‌تواند برای این باندها بهینه‌سازی شود.
- ارزیابی تنوع جمعیتی[2]: ارزیابی مدل روی پوست‌های روشن تا تیره ضروری است.
- طراحی رابط کاربری بالینی و ابزارهای تفسیرپذیری: برای پذیرش پزشکان، مدل باید نه‌تنها دقیق، بلکه قابل توضیح باشد.

در این پژوهش یک چارچوب نوین برای تشخیص ضایعات پوستی ارائه گردید که تلفیقی از فراسطح تصویربرداری چندطیفی و معماری یادگیری عمیق هیبریدی است. فراسطح طراحی‌شده قادر بود اطلاعات طیفی

1. multimodal
2. skin tone bias

غنی‌تری نسبت به تصاویر مرئی معمول فراهم آورد و مدل هیبریدی با بهره‌گیری از این داده‌ها توانست دقت، حساسیت و ویژگی بسیار بالایی در طبقه‌بندی ضایعات به دست آورد.

قدرت این روش در موارد زیر خلاصه می‌شود:

- استخراج اطلاعات طیفی حساس به ساختارهای پنهان ضایعه؛
- تحلیل هم‌زمان ویژگی‌های محلی و جهانی با معماری CNN–ViT؛
- عملکرد بالاتر نسبت به مدل‌های پیشین (دقت ~۹۸٪)؛
- تفسیرپذیری مناسب از طریق تحلیل نقشه‌های توجه؛
- قابلیت بالقوه برای ساخت ابزارهای قابل حمل و غیرتهاجمی در محیط بالینی.

این نتایج نشان می‌دهد ترکیب فناوری متاسرفیس با یادگیری عمیق می‌تواند مسیر جدیدی را در تشخیص پزشکی باز کند و زمینه‌ساز توسعه‌ی نسل آینده‌ی ابزارهای دقیق، سریع و کم‌هزینه برای تشخیص سرطان پوست باشد.

## منابع